\documentclass[aps,prl,twocolumn,superscriptaddress]{revtex4-2}
\usepackage[utf8]{inputenc}
\usepackage{graphicx}
\usepackage{hyperref}
\usepackage{amsmath}
\usepackage{todonotes}
\usepackage{physics}
\usepackage{siunitx}
\usepackage[version=4]{mhchem}
\usepackage{csquotes}

\begin{document}

\newcommand{\PVint}{\mathrm{p.v.}\int}
\newcommand{\step}{\theta}
\newcommand{\tens}[1]{\overset{\leftrightarrow}{\mathrm{#1}}}
\newcommand{\vecl}[1]{\overset{\leftarrow}{#1}}
\newcommand{\vecr}[1]{\overset{\rightarrow}{#1}}
\newcommand{\fefs}{\ce{^{57}Fe}}

\title{Collective nuclear excitation and pulse propagation in single-mode x-ray waveguides}
\author{Leon M. Lohse}
\email{llohse@uni-goettingen.de}
\affiliation{Georg-August-Universität Göttingen, 37077 Göttingen, Germany}
\affiliation{The Hamburg Centre for Ultrafast Imaging, 22761 Hamburg, Germany}
\affiliation{Deutsches Elektronen-Synchrotron DESY, 22607 Hamburg, Germany}

\author{Petar Andrejić}
\affiliation{Friedrich-Alexander-Universität Erlangen-Nürnberg, 91058 Erlangen, Germany}

\author{Sven Velten}
\affiliation{The Hamburg Centre for Ultrafast Imaging, 22761 Hamburg, Germany}
\affiliation{Deutsches Elektronen-Synchrotron DESY, 22607 Hamburg, Germany}

\author{Malte Vassholz}
\affiliation{Georg-August-Universität Göttingen, 37077 Göttingen, Germany}

\author{Charlotte Neuhaus}
\affiliation{Georg-August-Universität Göttingen, 37077 Göttingen, Germany}

\author{Ankita Negi}
\affiliation{The Hamburg Centre for Ultrafast Imaging, 22761 Hamburg, Germany}
\affiliation{Deutsches Elektronen-Synchrotron DESY, 22607 Hamburg, Germany}

\author{Anjali Panchwanee}
\affiliation{Deutsches Elektronen-Synchrotron DESY, 22607 Hamburg, Germany}

\author{Ilya Sergeev}
\affiliation{Deutsches Elektronen-Synchrotron DESY, 22607 Hamburg, Germany}

\author{Adriana Pálffy}
\affiliation{Julius-Maximilians-Universität Würzburg, 97074 Würzburg, Germany }

\author{Tim Salditt}
\affiliation{Georg-August-Universität Göttingen, 37077 Göttingen, Germany}

\author{Ralf Röhlsberger}
\affiliation{Friedrich-Schiller-Universität Jena, 07743 Jena, Germany}
\affiliation{Helmholtz-Institut Jena, 07743 Jena, Germany }
\affiliation{GSI Helmholtzzentrum für Schwerionenforschung GmbH, 64291 Darmstadt, Germany}
\affiliation{The Hamburg Centre for Ultrafast Imaging, 22761 Hamburg, Germany}
\affiliation{Deutsches Elektronen-Synchrotron DESY, 22607 Hamburg, Germany}

\date{\today}

\begin{abstract}
Waveguides offer a means to controllably couple atomic ensembles to the electromagnetic field therein. Here, we demonstrate x-ray propagation in planar thin-film waveguides coupled to Mössbauer nuclei under collective resonant excitation by short pulses of synchrotron radiation. We record x-ray photons that have been emitted into resonant modes of the waveguide. Depending on the geometry and mode of excitation, two fundamentally different signatures of the collective emission are observed, for which we present a unifying theoretical model. Our results form a new platform for waveguide quantum electrodynamics in the hard x-ray regime with the potential to provide a coherent narrowband source of x-rays on the nanometer scale.
\end{abstract}

\maketitle

When interacting with radiation, 
ensembles of identical quantum optical systems exhibit  collective dynamics  which is fundamentally different from that of the individual constituents \cite{Dicke_PR_1954}. Collective effects such as directed emission \cite{Scully_PRL_2006,Eberly_JoPBAMaOP_2006}, collective frequency shifts, superradiant decay \cite{Scully_PRL_2009}, and collective oscillations emerge already in the regime of weak excitation from as little as a single photon \cite{Scully_S_2009}.  Furthermore, in spatially extended ensembles, the geometry and spatial distribution distinctively affects the dynamics \cite{Svidzinsky_PRA_2010,Manassah_AiOaP_2012}. These effects have  been observed over various platforms, including one-dimensional chains of atoms coupled to optical waveguides
\cite{AsenjoGarcia_PRA_2017,Solano_NC_2017,Kumlin_PRA_2020,Pennetta_PRL_2022,CardenasLopez_PRR_2023,Su_PRR_2023} and Mössbauer nuclei in solid-state systems driven by hard x-rays \cite{Afanasev_JL_1965,Gerdau_PRL_1985,Buerck_PRL_1987,Shvydko_JoPCM_1989,Hannon_HI_1999,Smirnov_HI_1999,Buerck_HI_1999,Roehlsberger_PRL_2005,Roehlsberger_S_2010,Haber_NP_2017,Chumakov_NP_2017}. The latter are 
a particularly well-suited platform to observe and harness collective effects, 
since nuclear transitions resonant to hard x-ray frequencies have extremely narrow linewidths \cite{Roehlsberger_STiMP_2005}, rendering exceptionally clean quantum-optical systems \cite{Roehlsberger__2021,Adams_JoMO_2009}, such that Mössbauer nuclei were among the earliest systems to investigate collective radiative phenoma (see Refs. \cite{Afanasev_JL_1965,Kagan_JoPCSSP_1979,Hannon_HI_1999,Smirnov_HI_1999} and the references therein).


The experimental platforms available so far for quantum optics with Mössbauer nuclei have been either nuclear forward scattering (NFS) from unstructured foils \cite{Buerck_HI_1999}, nuclear Bragg diffraction (NBD) from crystals \cite{Afanasev_JL_1965,Gerdau_PRL_1985,Shvydko_JoPCM_1989,Chumakov_NP_2017}, or grazing-incidence reflection from thin-film structures \cite{Roehlsberger__2021}. Depending on the  experimental geometry, the collective nuclear excitation---also known as nuclear exciton \cite{Afanasev_JL_1965,Hannon_HI_1999,Smirnov_HI_1999}---has a different temporal dynamics. In NFS, i.e., a  thick slab or foil illuminated in normal incidence, the nuclear exciton exhibits so-called dynamical beats and emits into forward direction  \cite{Buerck_HI_1999,Smirnov_PRA_2007}. In contrast, a thin film embedded in an x-ray waveguide (WG) and excited in grazing-incidence illumination  decays exponentially and emits into the direction of specular reflection---which we call reflection geometry---with superradiant speedup and a shifted resonance frequency compared to the single atom \cite{Roehlsberger_S_2010}, similar to NBD \cite{Afanasev_JL_1965,Kagan_JoPCSSP_1979,Shvydko_JoPCM_1989}.
Still unexplored is a waveguide geometry, where identical nuclei are coupled to photons that propagate confined in a waveguide. The waveguide allows some control over the modes of the photon field and their dispersion. Such systems have attracted a lot of attention recently in the emerging field of waveguide QED \cite{Sheremet_RoMP_2023}, yet so far restricted to longer photon wavelengths.

\begin{figure}[htb]
    \centering
    \includegraphics{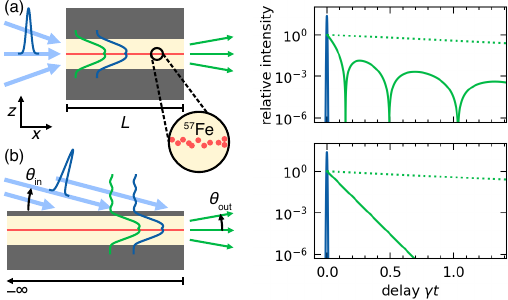}
    \caption{A short excitation pulse (blue) couples into a planar WG and creates a nuclear exciton in the \fefs\ layer (red) that radiatively decays (green). (a) FC excitation creates an exciton that emits pronounced temporal beatings. (b) RBC excitation of a long WG: the exciton decays exponentially. Superradiant speedup and shifted resonance, compared to the natural exponential decay with rate $\gamma$ (dotted green line), depend on the incidence angle $\theta_\mathrm{in}$. The plots on the right show exemplary theory calculations.  }
    \label{fig:geometries}
\end{figure}

In this Letter, we demonstrate the excitation and collective dynamics of Mössbauer nuclei coupled in forward incidence to a single-mode nanometer-thin x-ray waveguide, resolved in both time and position. We present two experiments, in which we excite the sample either in front-coupling (FC) geometry [Fig.\ \ref{fig:geometries}(a)] \cite{Fuhse_APL_2004} or resonant-beam-coupling (RBC) geometry [Fig.\ \ref{fig:geometries}(b)] \cite{Feng_PRL_1993}. 
Our experimental technique combined with optimized WG designs and fabrication allows to observe the photons that have been emitted into resonant modes of the WG leaving its back end.
Depending on the way of excitation, the decay of the nuclear exciton either exhibits dynamical beats as in NFS or it decays exponentially with superradiant speedup. We explain our observations by a unifying theoretical model based on a recently introduced approach based on a real-space Green’s function formalism \cite{Andrejic_PRA_2024}. 
Our results shed new light on earlier experiments in reflection geometry and open new ways to design spatio-temporal properties of the nuclear exciton based on hard x-ray waveguiding, thus benefiting the emerging field of waveguide QED \cite{Sheremet_RoMP_2023}. 

Hard x-rays are an extreme regime for waveguiding.
In the energy regime above \SI{10}{\keV}, refractive indices are written as $n= 1- \delta + i \beta$, where $\delta, \beta < \num{e-5}$. X-ray waveguiding manifests when lighter core materials are surrounded by denser cladding materials.
Because of the small contrasts in $n$, x-ray WGs are weakly guiding (modes typically extend over a few 100 wavelengths in the transversal direction) and cause significant photo-electric absorption in the dense cladding. 
Consequently, the Purcell enhancement is negligible and coupling to resonant modes is small \cite{Lohse_OE_2024}.
However, dense cladding strongly attenuates everything but the lowest few resonant modes \cite{Osterhoff_NJoP_2011}, so that fields generated by emitters in the WG can be asymptotically approximated by only their resonant components \cite{Lohse_OE_2024}.

We have prepared two planar x-ray WGs, both containing a few-atomic-layers-thin iron film, to \SI{95}{\percent} enriched in \fefs, in the center of its guiding layer (see Supplemental Material \cite{supp}). The \fefs\ does not develop long-range magnetic order in this thin film and hence exhibits essentially a single resonance line.
The experiments were performed at the dynamics beamline P01 at PETRA III (DESY, Hamburg). 
The samples were illuminated by \SI{100}{\ps}-long synchrotron pulses with photon energy \SI{14.4}{\keV} (monochromatized to \SI{1}{\meV} bandwidth) at a repetition period of \SI{192}{\ns}, creating nuclear excitons. The photons leaving the WG at its back end were detected with a stack of avalanche photo diodes (APDs). By temporal gating, the (delayed) emissions of the nuclear exciton are detected independently from the (prompt) synchrotron pulse.

\begin{figure}[htb]
    \centering
    \includegraphics{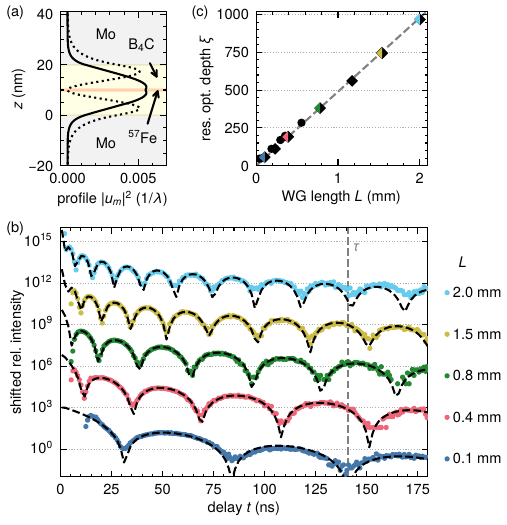}
    \caption{Layer design and experimental data for FC excitation. (a) Transversal mode profiles $u_m(z)$ of the 2 guided modes supported by the WG. (b) Experimentally observed emissions as a function of delay $t$ after excitation for different WG lengths $L$ (colored points). The intensities are normalized and vertically shifted. The dashed lines show simulations of NFS through \fefs-foils, using hyperfine parameters and resonant optical depth $\xi$ that were estimated from the experimental data. (c) Extracted resonant optical depth $\xi$ for different $L$ (diamonds and discs indicate 2 independent measurements).}
    \label{fig:expfronttime}
\end{figure}

First, we have prepared a layer system optimized for FC excitation, consisting of a \SI{0.6}{\nm}-thin layer of \fefs, embedded in a planar x-ray WG with \SI{20}{\nm} \ce{B_4C} core, and symmetric \SI{30}{\nm}-thick \ce{Mo} cladding as sketched in Fig.\ \ref{fig:expfronttime}(a).
The layer structure was deposited onto a \SI{1}{\mm}-thick Ge wafer and capped with a second Ge wafer after deposition as detailed in Ref.\ \cite{Krueger_JoSR_2012}. 
The synchrotron beam was focussed with two elliptical mirrors in Kirkpatrick-Baez geometry to a spot of about \SI{7}{\um} diameter into the WG entrance at the focal position.
The Ge wafers were used to absorb the tails of the focus to reduce background signal, but they do not affect the resonant mode structure due to the thick cladding \cite{Salditt_PRL_2008}. The wafer sandwich was cut to triangular shape, allowing us to change the effective WG length $L$ by translating it along $y$ transversal to the beam (see Supplemental Material \cite{supp}). The WG length was calibrated by measuring the absorption in the wafer as a function of $y$-translation.
Figure \ref{fig:expfronttime}(b) shows the temporal evolution of the emitted x-rays for several WG lengths up to \SI{2}{\mm} (limited by off-resonant absorption). The emissions clearly exhibit dynamical beats resembling those known from NFS through a homogeneous single-line resonant absorber. 
The dashed lines show simulations of NFS through a thick foil of enriched \fefs\, including inhomogeneous hyperfine-splitting in the order of $6$ natural linewidths, in a 4-parameter model (see Supplemental Material \cite{supp}), which accurately describes the data. The optimal parameters were found using a maximum-likelihood estimation assuming Poisson statistics. The magnitude of hyperfine splitting is expected from similar amorphous thin films \cite{Roehlsberger_S_2010,Sahoo__2011}.
The extracted optical depth $\xi$ (also called \enquote{effective thickness} \cite{Buerck_HI_1999}) [Fig.\ \ref{fig:expfronttime}(c)] is strictly proportional to the WG length $L$.

\begin{figure}[tbh]
    \centering
    \includegraphics{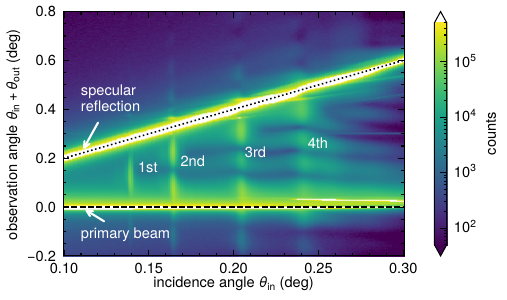}
    \caption{Off-resonant intensity of the RBC sample, illuminated at an incidence angle of $\theta_\mathrm{in}$ and recorded \SI{3}{\metre} downstream of the sample with a pixel detector. Vertical cuts correspond to observed far-field patterns. The 4 resonant modes, which are supported by the layer structure, are clearly visible in the region between the primary beam ($\theta_\mathrm{in} + \theta_\mathrm{out} = 0$) and the specularly reflected beam ($\theta_\mathrm{out} = \theta_\mathrm{in}$). The modes are well separated in $\theta_\mathrm{in}$.}
    \label{fig:expmodes}
\end{figure}

\begin{figure}[tbh]
    \centering
    \includegraphics{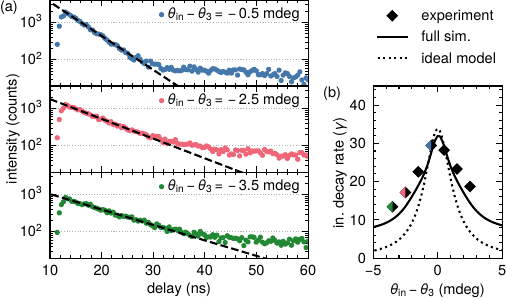}
    \caption{Experimentally measured emissions of the nuclear exciton in RBC excitation. (a) Decay pattern for 3 angles in the vicinity of the 3rd resonant mode and initial exponential decay (black dashed line). (b) Initial decay rate as a function of angle detuning: extracted from the experimental data (black diamonds); calculated with the simplified ideal model \eqref{eq:speedup} (dashed line); simulation including  hyperfine splitting of the transition energies as well as the angular divergence of the experiment (solid line).}
    \label{fig:expgitime}
\end{figure}

For the second experiment, we have prepared a layer system optimized for RBC excitation [Fig.\ \ref{fig:geometries}(b)], having a \SI{40}{\nm}-thick \ce{B_4C} core and, in contrast to the FC experiment, a top Mo cladding layer of only \SI{8}{\nm} to allow incident light to evanescently couple into the WG. We used a \SI{10}{\mm} long Si wafer as substrate, so that the WG is significantly longer than the attenuation length of the resonant modes and thus has effectively infinite extent. The back end of the sample was broken off after deposition to create a clean exit face. The sample was placed onto a goniometer stage for tuning the incidence angle $\theta_\mathrm{in}$.
Figure \ref{fig:expmodes} shows the off-resonant far-field intensity distribution, detected \SI{3}{\metre} downstream of the sample by a time-insensitive pixel detector as a function of $\theta_\mathrm{in}$.  
We then aligned the APDs with the WG exit [$\theta_\mathrm{out} \approx \ang{0}$, see Fig.\ \ref{fig:geometries}(b)] and collected the emitted photons for several values of $\theta_\mathrm{in}$ in the vicinity of the third resonance $\theta_3 = \ang{0.203}$. 
We selected the 3rd mode for practical reasons, because the divergence of the incident beam was \ang{0.002}, which is wider than the resonance of the 1st mode. Figure \ref{fig:expgitime}(a) shows the emitted intensity as a function of time for three different $\theta_\mathrm{in}$. The intensity exhibits an initial superradiant decay and a slow-down for longer delays, which is caused by a residual hyperfine splitting of the nuclear levels. The initial decay is plotted in panel (b), showing a decay rate of $30 \gamma$ at $\theta_3$, which gradually diminishes as $\theta_\mathrm{in}$ moves away from resonance. Similar behaviour was previously observed with NBD \cite{Shvydko_JoPCM_1989} and thin films in reflection geometry \cite{Roehlsberger_S_2010}. The solid line shows a simulation in reflection geometry \cite{supp} that considers divergence of the incoming beam and hyperfine splitting of the transitions. 
The hyperfine parameters for the simulation were extracted from measurements at $\theta_\mathrm{in} = \ang{0.5}$, far from the WG modes, and are consistent with the FC experiment \cite{supp}.

To qualitatively model the experimental observations, we employ the theory developed in Ref.\  \cite{Andrejic_PRA_2024} based on macroscopic quantum electrodynamics, which describes the atom-light interaction using the classical electromagnetic Green's function.
Importantly, it explicitly models the propagating field modes in the dispersive waveguide environment with spatial dependency, which has not been relevant for the existing theories of NFS \cite{Smirnov_PRA_2007,Shvydko_PRB_1999}, NBD \cite{Hannon_HI_1999,Kagan_JoPCSSP_1979}, or the reflection geometry \cite{Heeg_PRA_2013,Lentrodt_PRR_2020,Lentrodt_PRL_2023}.  
Here, we neglect the nearly degenerate sublevels of the ground ($I_g = 1/2$) and excited ($I_e=3/2$ ) state of \fefs\ ($\hbar \omega_0 = \SI{14.4}{\keV}$, $\hbar \gamma = \SI{4.7}{\nano\eV}$, M1-transition) and describe the nuclear state by a single nuclear transition operator for each atom $\hat \sigma^i_{ge} = \dyad{g}{e}$. 
The excitation pulses have very small pulse areas so that the nuclear excited states remain unpopulated ($\ev{\hat \sigma^i_{ee}} \approx 0$) and the response is linear. 
The expectation values of the transition operators $\sigma^i_{ge} := \ev{\hat \sigma^i_{ge}}$ evolve, in the rotating frame of the nuclear transition frequency $\omega_0$, according to \cite{AsenjoGarcia_PRA_2017}
\begin{equation}
\begin{split}
\label{eq:dynamics1}
    -i \omega \sigma^i_{ge}(\omega) 
    &= - \frac{\gamma}{2} \sigma^i_{ge}(\omega) + i \Omega(\vb r_i, \omega) \\
    &+ i \sum_{j \neq i}  g_{ij}(\omega) \sigma^j_{ge}(\omega),
\end{split}
\end{equation}
where $\Omega(\vb r, t) = \vb m_0^* \cdot \vb B_\mathrm{in}(\vb r, t) / \hbar$ is the propagating excitation pulse expressed as a Rabi frequency, $\vb m_0$ the magnetic dipole moment of the transition and $g_{ij}$ are dispersive coupling coefficients.
The total single-atom spontaneous decay rate $\gamma$ includes self-interaction and is equal to its value in a homogeneous environment (negligible Purcell factor). Note that $\vb B_\mathrm{in}$ is the free incident magnetic field and scattering from the nuclei is fully accounted for by $g_{ij}$.
The coupling is mediated by the WG environment via
\begin{equation}
\label{eq:coupling1}
    g_{ij}(\omega) := 
    \frac{\mu_0 k_0^2}{\hbar} \, \vb m_0^* \cdot \tens{G}_{\mathrm{m}}(\vb r_i, \vb r_j, \omega_0 + \omega) \cdot \vb m_0,
\end{equation}
where $\mu_0$ is the vacuum permeability, $k_0 = \omega_0 / c$, and $\tens{G}_{\mathrm{m}}$ the Green's function for the magnetic field in the WG (see Supplemental Material \cite{supp}). While the $\sigma_{ge}^i$ are defined in the rotating frame, this Green's function depends on the absolute frequency $\omega_0 + \omega$. 

We convert \eqref{eq:dynamics1} into a one-dimensional macroscopic equation. To that end, we assume homogeneously distributed nuclei in a single thin layer, insert the analytic expression for the Green's function, and assume interaction via a single resonant mode. We obtain
\begin{multline}
\label{eq:dynamics2}
    \dot \sigma_{ge}(x, t) 
    = - \frac{\gamma}{2} \sigma_{ge}(x, t) + i \Omega(x, t) \\
    - \frac{\zeta_m \gamma}{4 \Lambda_\mathrm{res}}\int_{-L}^0 \dd x' e^{i k_0 \nu_m\abs{x -x'}} \sigma_{ge}(x', t - \abs{x - x'} \nu_m / c).
\end{multline}
Here, $\Lambda_\mathrm{res}$ is the on-resonance attenuation length (\SI{61}{\nm} for bulk \fefs), $\nu_m$ is the complex effective refractive index of the mode \cite{Lohse_OE_2024}, and $\zeta_m$ the dimensionless coupling-coefficient of the WG mode.  
The latter effectively rescales $\Lambda_\mathrm{res}$ and is given by $\zeta_m = d u_m(z_0)^2$, where $d$ and $z_0$ are thickness and position of the \fefs-layer, respectively and $u_m(z)$ is the complex transversal mode profile, described in Ref.\ \cite{Lohse_OE_2024}, which is bi-normalized to $\int [u_m(z)]^2 \dd z = 1$. 
The field radiated by the nuclear exciton is obtained via
\begin{equation}
\vb B^+_\mathrm{nuc}(\vb r, \omega) = \mu_0 k_0^2 \sum_j \tens{G}_{\mathrm{m}}(\vb r, \vb r_j, \omega_0 + \omega) \cdot \vb m_0 \sigma_{ge}^j(\omega),
\end{equation}
corresponding to the field radiated by the oscillating magnetization $\vb M = \vb m_0 \rho \sigma_{ge} \exp(-i \omega_0 t) + \text{h.c.}$.
Note that \eqref{eq:dynamics2} not only describes planar WGs (after integrating out $y$) as discussed here, but also one-dimensional channel WGs, and furthermore homogeneous slabs (foils) with translational symmetry in $y$-$z$ \cite{Friedberg_PLA_2008a}, with different values for $\nu_m$ ($=n$ for slab) and $\zeta_m$ ($=1$ for slab). 

Equation \eqref{eq:dynamics1} is often treated as an eigenvalue problem,  decomposing $\sigma_{ge}$ into radiative eigenmodes \cite{AsenjoGarcia_PRA_2017,Svidzinsky_PRA_2010,Manassah_AiOaP_2012}. Instead, we employ a forward-scattering approximation that allows us to solve \eqref{eq:dynamics2} analytically for the two experimental settings sketched in Fig.\ \ref{fig:geometries}. The rapidly oscillating phase factor in \eqref{eq:dynamics2} strongly suppresses the back-scattered field \cite{Andrejic_PRA_2024}. Hence, we change the upper limit of integration in \eqref{eq:dynamics2} to $x$.

In the FC geometry [Fig.\ \ref{fig:geometries}(a)], the only surviving component of the excitation pulse is the fundamental guided mode, so that $\Omega(x,t) = \Omega_0 \exp(i k_0 \nu_m  x) \Pi(t_\mathrm{ret})$, with some slowly varying envelope $\Pi(t)$ where $t_\mathrm{ret} = t - k_0 \nu_m x / c$. Note that the excitation pulse decays with $x$, as $\Im\{\nu_m\} > 0$. 
For finite $L$, \eqref{eq:dynamics2} becomes a Volterra integral equation due to the forward-scattering approximation and hence has no radiative eigenmodes. Its exact analytical solution for the given $\Omega(x, t)$ is 
\begin{equation}
\label{eq:sigmafront}
    \sigma_{ge}(x, t)  =i A e^{i k_0 \nu_m x -\gamma t_\mathrm{ret}/2} J_0\left[\sqrt{\gamma t_\mathrm{ret} x \zeta_m \Lambda^{-1}_\mathrm{res}}\right],
\end{equation}
for $t_\mathrm{ret} \ge 0$,
where $J_0$ is a Bessel function of the first kind and  $A = \Omega_0 \int \Pi(t) \dd t$ the pulse area.
This is the well known solution for NFS \cite{Shvydko_PRB_1999,Smirnov_PRA_2007} with $\Lambda_\mathrm{res}$ rescaled by $\zeta_m$. 
Computing the emitted field from \eqref{eq:sigmafront} is straight-forward (see Supplemental Material \cite{supp}). The exciton emits directionally and purely into the WG mode. The emission undergoes the dynamical beats, well known from coherent pulse propagation through resonant media \cite{Kagan_JoPCSSP_1979,Buerck_HI_1999}---and clearly observed in our experiment. 
From the experimental data, we obtain an estimate for the coupling coefficient $\abs{\zeta_m} = \xi \Lambda_\mathrm{res} / L \approx \num{0.029}$, slightly smaller than the theoretical value (from design parameters) of \num{0.038}.

In the RBC geometry [Fig.\ \ref{fig:geometries}(b)], the incidence angle $\theta_\mathrm{in}$ fixes the in-plane wavevector to $k_\mathrm{in} = k_0 \cos \theta_\mathrm{in}$.
The excitation pulse can be written as $\Omega(x,t) = \Omega_0 \exp(i k_\mathrm{in} x) \Pi(t_\mathrm{ret})$, with here $t_\mathrm{ret} = t - k_\mathrm{in} x / \omega_0$. 
Assuming that $L$ is significantly larger than the off-resonance attenuation length of the mode, $\Lambda_m = (2 k_0 \Im\{\nu_m\})^{-1}$, (typically a few \SI{100}{\um}) we can extend the lower boundary of integration in \eqref{eq:dynamics2} to $-\infty$.
This eliminates the boundary effect in the forward direction. 
The linear spatial phase in $\Omega(x,t)$ corresponds to a radiative eigenmode of the exciton, so that the exciton evolves harmonically with shifted frequency and enhanced decay. 
We obtain (see Supplemental Material \cite{supp})
\begin{align}
\label{eq:expdecay}
\sigma_{ge}(x, t) 
    &= i A \exp\left[ik_0x \cos\theta_\mathrm{in} - \left(\frac{\gamma }{2} - i \eta \right)t_\mathrm{ret}\right], \\
    \label{eq:speedup}
    \eta(\theta_\mathrm{in}) 
    &=   \frac{\zeta_m \Lambda_m }{\Lambda_\mathrm{res}} \frac{1}{2 \Lambda_m q_\mathrm{\theta} - i}\frac{\gamma}{2},
\end{align}
with $q_\mathrm{\theta}  = k_0 \cos \theta_\mathrm{in} - k_0 \Re\{ \nu_m\} \approx - k_0 (\theta_\mathrm{in} - \theta_m)\sin \theta_m $ and pulse area $A$ as previously defined. The mode angle is defined via $\cos \theta_m =\Re\{ \nu_m\}$. Similar to NBD \cite{Kagan_JoPCSSP_1979,Shvydko_JoPCM_1989}, the exciton decays superradiantly with peak speedup of $1 + \zeta_m \Lambda_m / \Lambda_\mathrm{res}$ depending on ratio between off- and on-resonant attenuation lengths, as well as the mode coupling coefficient $\zeta_m$.
Note that $\Lambda_m / \Lambda_\mathrm{res} = \rho \Lambda_m \sigma_\mathrm{res}$ ($\sigma_\mathrm{res}$: resonant cross section, $\rho$: number density of nuclei) gives the number of nuclei within the off-resonant absorption length (compare Ref. \cite{Adams_JoMO_2009}), clearly demonstrating the collective superradiant nature of the speedup, contrasting earlier interpretations based on an increase of the photon density of states \cite{Roehlsberger_PRL_2005}.
Figure \ref{fig:expgitime}(b) shows the speedup from calculations (dashed line) with \eqref{eq:speedup} using the design parameters as well as the speedup from the experimental data (diamonds). This simple single-mode model already explains the observed peak speedup. It does not fully account for the increased angular resonance width, which can be attributed to the residual hyperfine splitting and angular divergence observed in the experiment \cite{supp}. Thus, we see that \eqref{eq:dynamics2} qualitatively reproduces both experiments. Notably, similar eigenmodes \eqref{eq:expdecay} are observed in crystals \cite{Shvydko_JoPCM_1989} and waveguides in grazing-incidence reflection geometry \cite{Roehlsberger__2021}.

In summary, we have demonstrated the excitation and collective dynamics of Mössbauer nuclei in x-ray waveguides for two different modes of excitation. 
In the FC geometry, we have observed dynamical beats emitted by WGs with a \SI{20}{\nm}-thick guiding core and varying lengths from about \SI{0.1}{\mm} to \SI{2}{\mm} into the fundamental WG mode. In the RBC geometry, we have observed exponential decay into the resonantly excited WG mode with tuneable superradiant speedup.
Our theory consistently describes the observations of both experiments in a one-dimensional waveguide picture.
When a single mode dominates, the equation of motion is formally identical to that of a homogeneous slab with length (thickness) rescaled by a nucleus-mode coupling constant $\zeta_m$. Depending on the off-resonant optical thickness of the WG, two regimes are realized.
For optically thin WGs, the finite integrated nuclear density gives rise to dynamical beats.
The grazing-incidence geometry allows the study of optically thick WGs, where off-resonant absorption suppresses boundary effects, establishing translation invariance and permitting radiative eigenmodes, so that the exciton evolves harmonically with resonant dispersion relation.

Our results open up a number of promising future directions.
We demonstrated how to ``harvest'' the propagating beam at the exit of the WG, forming an extremely coherent narrowband source of \SI{20}{\nm} cross section, albeit with low brightness. Both the spatial diffraction patterns and temporal dynamics can be readily resolved at these scales.
The FC geometry allows to simultaneously and coherently couple into multiple guided modes in one or several WGs.
A multi-mode WG could be used to implement effective two-beam control techniques for the x-ray frequency regime. 
The spatial coherence allows to spatially separate individual modes and permits new approaches for narrowband x-ray control via the engineering of collective radiation patterns. 
The source may be utilized for investigations at the nanoscale that leverage both spatial and temporal coherence.

\begin{acknowledgments}
   We thank Mike Kanbach for preparing the WG assemblies, Olaf Leupold for his support and fruitful discussions during the first beamtime, and Dieter Lott and Helmholtz-Zentrum Hereon for providing and supporting the reflectometer to calibrate the thin film samples. We acknowledge DESY (Hamburg, Germany), a member of the Helmholtz Association HGF, for the provision of experimental facilities. Beamtime was allocated for proposals I-20211677; I-20221193. We acknowledge partial funding by Max Planck School of Photonics; Deutsche Forschungsgemeinschaft (DFG) (432680300 SFB 1456/C03, 429529648 TRR 306/C04, 390858490 EXC 2147, 390715994 EXC 2056). A.~Pálffy acknowledges support from the DFG in the framework of the Heisenberg Program.
   LML, MV, TS, and RR are part of the Max Planck School of Photonics.
\end{acknowledgments}

\bibliography{references}

\end{document}


\newcommand{\PVint}{\mathrm{p.v.}\int}
\newcommand{\step}{\theta}
\newcommand{\tens}[1]{\overset{\leftrightarrow}{\mathrm{#1}}}
\newcommand{\vecl}[1]{\overset{\leftarrow}{#1}}
\newcommand{\vecr}[1]{\overset{\rightarrow}{#1}}

\newcommand{\fefs}{\ce{^{57}Fe}}

\DeclareSIUnit\bar{bar}

\setcounter{equation}{0}
\setcounter{figure}{0}
\setcounter{table}{0}
\setcounter{page}{1}
\makeatletter
\renewcommand{\theequation}{S\arabic{equation}}
\renewcommand{\thefigure}{S\arabic{figure}}
\renewcommand{\thetable}{S\arabic{table}}
\renewcommand{\bibnumfmt}[1]{[S#1]}
\renewcommand{\citenumfont}[1]{S#1}

\title{Supplemental Material to: Collective nuclear excitation and pulse propagation in single-mode x-ray waveguides}
\author{Leon M. Lohse}
\email{llohse@uni-goettingen.de}
\affiliation{Georg-August-Universität Göttingen, 37077 Göttingen, Germany}
\affiliation{The Hamburg Centre for Ultrafast Imaging, 22761 Hamburg, Germany}
\affiliation{Deutsches Elektronen-Synchrotron DESY, 22607 Hamburg, Germany}

\author{Petar Andrejić}
\affiliation{Friedrich-Alexander-Universität Erlangen-Nürnberg, 91058 Erlangen, Germany}

\author{Sven Velten}
\affiliation{The Hamburg Centre for Ultrafast Imaging, 22761 Hamburg, Germany}
\affiliation{Deutsches Elektronen-Synchrotron DESY, 22607 Hamburg, Germany}

\author{Malte Vassholz}
\affiliation{Georg-August-Universität Göttingen, 37077 Göttingen, Germany}

\author{Charlotte Neuhaus}
\affiliation{Georg-August-Universität Göttingen, 37077 Göttingen, Germany}

\author{Ankita Negi}
\affiliation{The Hamburg Centre for Ultrafast Imaging, 22761 Hamburg, Germany}
\affiliation{Deutsches Elektronen-Synchrotron DESY, 22607 Hamburg, Germany}

\author{Anjali Panchwanee}
\affiliation{Deutsches Elektronen-Synchrotron DESY, 22607 Hamburg, Germany}

\author{Ilya Sergeev}
\affiliation{Deutsches Elektronen-Synchrotron DESY, 22607 Hamburg, Germany}

\author{Adriana Pálffy}
\affiliation{Julius-Maximilians-Universität Würzburg, 97074 Würzburg, Germany }

\author{Tim Salditt}
\affiliation{Georg-August-Universität Göttingen, 37077 Göttingen, Germany}

\author{Ralf Röhlsberger}
\affiliation{Friedrich-Schiller-Universität Jena, 07743 Jena, Germany}
\affiliation{Helmholtz-Institut Jena, 07743 Jena, Germany }
\affiliation{GSI Helmholtzzentrum für Schwerionenforschung GmbH, 64291 Darmstadt, Germany}
\affiliation{The Hamburg Centre for Ultrafast Imaging, 22761 Hamburg, Germany}
\affiliation{Deutsches Elektronen-Synchrotron DESY, 22607 Hamburg, Germany}

\maketitle

\section{Theory}
\subsection{Detailed derivation of the equation of motion}

We derive the equations of motion from the main manuscript using the approach detailed in Ref.\ \cite{Andrejic_PRA_2024}. It is based on the Gruner-Welsch quantization of the macroscopic Maxwell's equations \cite{Gruner_PRA_1996}, which expresses the electromagnetic field in terms of the classical dyadic Green's function of the medium. We use an asymptotic expansion of the Green's function in terms of its resonant (guided and leaky) and non-resonant modes, derived in Ref.\ \cite{Lohse_OE_2024}, to obtain tractable, yet quantitative, analytical expressions for the coupling terms. Refer to the references given in the introductions of Refs. \cite{Andrejic_PRA_2024,Lohse_OE_2024} for an extensive list of related works.

The \SI{14.4}{\keV} resonance of the \fefs\ nucleus corresponds to a magnetic dipole (M1) transition between one of the ground and excited states with nuclear spins $I_g = 1/2$ and $I_e = 3/2$, respectively. For the purpose of this letter, we neglect the $2 I + 1$ sublevels and consider only a single ground $\ket{g}$ and a single excited state $\ket{e}$, with a magnetic transition dipole moment that is aligned with the polarization of the excitation pulse. The full picture is discussed in Ref.\ \cite{Andrejic_PRA_2024}.
We write the nuclear transition operator, in the rotating frame of the nuclear transition frequency, as $\hat\sigma_{ge} := \op{g}{e}$, so that the magnetic dipole moment operator is
\begin{equation}
    \hat{\vb m}^i = \vb m_0 \pqty{\hat{\sigma}^i_{ge} e^{- i\omega_0 t}+ \hat{\sigma}^i_{eg} e^{ i \omega_0 t} },
\end{equation}
with magnetic dipole moment $\vb m_0$ and transition energy $\hbar \omega_0$. The diagonal operators for the excited and ground states are written accordingly. Similarly, we separate the magnetic field operators into components with positive and negative frequencies, such that
\begin{equation}
    \hat{\vb B}(t) = \hat{\vb B}^{+}(t) e^{-i \omega_0 t} + \hat{\vb B}^{-}(t) e^{i \omega_0 t}.
\end{equation}
The nuclear transition operators then fulfill \cite{Andrejic_PRA_2024}
\begin{equation}
\label{eq:smicroscopic0}
    \dv{t}\hat{\sigma}^i_{ge}(t) 
    = - \frac{\gamma}{2} \hat{\sigma}^i_{ge}(t) - \bqty{ \hat{\sigma}^i_{ee}(t) - \hat{\sigma}^i_{gg}(t) } \frac{i}{\hbar} \vb m_0^* \cdot \hat{\vb B}^{+}(\vb r_i, t),
\end{equation}
under the rotating wave approximation. Since the pulse areas of the synchrotron pulses are tiny, even under extreme focussing, the excited state populations will be very small and negligible in \eqref{eq:smicroscopic0}. We make the approximation $\bqty{ \hat{\sigma}^i_{ee}(t) - \hat{\sigma}^i_{gg}(t) } \approx -1$. Taking the expectation values on both sides yields
\begin{equation}
\label{eq:smicroscopic}
    \dot{\sigma}^i_{ge}(t) 
    = - \frac{\gamma}{2} \sigma^i_{ge}(t) + \frac{i}{\hbar} \vb m_0^* \cdot \vb B^{+}(\vb r_i, t).
\end{equation}
The field consists of the incoming field and the field emitted by the nuclei,
$\vb B = \vb B_\mathrm{in} + \vb B_\mathrm{nuc}$. The nuclei emit the field \cite{Andrejic_PRA_2024}
\begin{equation}
\label{eq:bmicroscopic}
\begin{split}
    \vb B^+_\mathrm{nuc}(\vb r, \omega) 
    &= \mu_0 \pqty{\frac{\omega+\omega_0}{c}}^2 \sum_i \tens{G}_\mathrm{m}(\vb{r}, \vb{r}_i, \omega+\omega_0) \cdot (\vb m^i)^+(\omega) \\
    &= \mu_0 \pqty{\frac{\omega+\omega_0}{c}}^2 \sum_i \tens{G}_\mathrm{m}(\vb{r}, \vb{r}_i, \omega+\omega_0) \cdot \vb m_0\,\sigma^i_{ge}(\omega),
\end{split}
\end{equation}
where 
\begin{equation}
    \Bqty{\epsilon(\vb r) \nabla \times \frac{1}{\epsilon(\vb r)} \nabla \times - [n(\vb r) \omega/c]^2 } \tens{G}_\mathrm{m}(\vb r, \vb r') = \epsilon(\vb r) \delta(\vb r - \vb r') \tens{1}
\end{equation}
defines the magnetic Green's function. Here $\epsilon$ is the relative electric permeability and $n$ the refractive index of the layer system, excluding the nuclear interaction.
Inserting \eqref{eq:bmicroscopic} into the Fourier transform of \eqref{eq:smicroscopic}, we obtain
\begin{equation}
    -i \omega\sigma^i_{ge}(\omega) 
    = - \frac{\gamma}{2} \sigma^i_{ge}(\omega) + i \Omega(\vb r_i, \omega) + i \sum_{j \neq i} g_{ij}(\omega+\omega_0) \sigma^j_{ge}(\omega),
\end{equation}
where $\Omega(\vb r, \omega) = \vb m_0^* \cdot \vb B^+_\mathrm{in}(\vec r, \omega) / \hbar$ is the time-dependent Rabi frequency produced by our propagating excitation pulse and coupling between the nuclei is mediated by
\begin{equation}
\label{eq:coupling1}
    g_{ij}(\omega) = 
    \frac{\mu_0 \omega^2}{\hbar c^2} \, \vb m_0^* \cdot \tens{G}_{\mathrm{m}}(\vb r_i, \vb r_j, \omega) \cdot \vb m_0.
\end{equation}

The waveguide contains a large number of nuclei that are homogeneously distributed in an amorphous thin layer at height $z = z_0$. Let $d$ denote the layer thickness and $\rho$ the nuclear number density within the layer.
Going to continuous variables and integrating out the transversal dimensions $y$ and $z$, we obtain
\begin{equation}
\label{eq:dynamics1}
    -i \omega\sigma_{ge}(x, \omega) 
    = - \frac{\gamma}{2} \sigma_{ge}(x, \omega) + i \Omega(x, z_0, \omega) + i \rho d\ \PVint g(x-x',\omega+\omega_0) \sigma_{ge}(x', \omega) \dd{x'},
\end{equation}
with
\begin{equation}
\label{eq:coupling2}
    g(x - x', \omega) = 
    \frac{\mu_0 \omega^2}{\hbar c^2} \, \vb m_0^* \cdot \tens{G}_{\mathrm{m}, \mathrm{2d}}(x-x', z_0, z_0, \omega) \cdot \vb m_0.
\end{equation}
where $\tens{G}_{\mathrm{m}, \mathrm{2d}}$ is the Green's function corresponding to a line source extended along $y$ (see Ref.\ \cite{Lohse_OE_2024} for a detailed discussion). For x-ray waveguides, it is approximately transversal and can be expanded into its resonant and non-resonant modes, giving
\begin{equation}
\label{eq:gfexpansion}
 \tens{G}_{\mathrm{m}, \mathrm{2d}}(x-x', z, z', \omega) \approx \tens{1}_\perp \frac{i}{\omega/c} \sum_m \frac{u_m(z) u_m(z')}{2 \nu_m} e^{i \omega \abs{x-x'}\nu_m /c} + \text{non-resonant},
\end{equation}
where $\nu_m$ are the (complex) effective refractive indices of the corresponding resosonant modes (in short: mode indices), $\tens{1}_\perp = \vu{y} \otimes \vu{y} + \vu{z} \otimes \vu{z}$ the transversal unit tensor, and $u_m(z)$ are transversal mode profiles as defined in Ref.\ \cite{Lohse_OE_2024}. They are bi-normalized according to
\begin{equation}
 \int \frac{[u_m(z)]^2}{\epsilon(z)}  \dd{z} = 1.
\end{equation}
The expansion \eqref{eq:gfexpansion} in particular uses the fact that transverse magnetic and transverse electric modes are approximately degenerate in the hard x-ray regime due to the small refractive index contrasts, so that the polarization-dependence simplifies to the transversal unit tensor.

We make the crucial approximation to neglect the non-resonant modes, since they are strongly attenuated, and obtain
\begin{equation}
\label{eq:dynamics2}
    - i \omega \sigma_{ge}(x, \omega) 
    = - \frac{\gamma}{2} \sigma_{ge}(x, \omega) + i \Omega(x, \omega) 
    - \frac{\mu_0 \abs{\vb m_0}^2 \rho}{2 \hbar} \frac{\omega+\omega_0}{c} \sum_m \zeta_m \int_{-L}^0 \dd{x'}  e^{i (\omega+\omega_0) \abs{x -x'}\nu_m /c} \sigma_{ge}(x', \omega).
\end{equation}
Since $\omega \ll \omega_0$, we neglect the frequency detuning in the prefactor.
Fourier-transforming \eqref{eq:dynamics2} finally yields
\begin{equation}
    \dot \sigma_{ge}(x, t) 
    = - \frac{\gamma}{2} \sigma_{ge}(x, t) + i \Omega(x, t) 
    - \frac{\gamma }{4 \Lambda_\mathrm{res}} \sum_m \zeta_m \int_{-L}^0 \dd{x'} e^{i k_0 \nu_m\abs{x -x'}} \sigma_{ge}(x', t - \abs{x - x'} \nu_m / c),
\end{equation}
where $k_0 = \omega_0 / c$ and
\begin{equation}
    \frac{1}{\Lambda_\mathrm{res}} = \rho \frac{k_0 \mu_0 \abs{\vb m_0}^2}{\hbar} \frac{2}{\gamma} \equiv \rho \sigma_\mathrm{res}
\end{equation}
is the reciprocal on-resonance attenuation length.

Making the forward-scattering approximation and assuming that only a single mode contributes gives
\begin{equation}
\label{eq:dynamicsfinal}
    \dot \sigma_{ge}(x, t) 
    = - \frac{\gamma}{2} \sigma_{ge}(x, t) + i \Omega(x, t) 
    - \frac{\gamma \zeta_m}{4 \Lambda_\mathrm{res}} \int_{-L}^x \dd{x'} e^{i k_0 \nu_m(x -x')} \sigma_{ge}\pqty{x', t - \nu_m \frac{x - x'}{c}}.
\end{equation}

\subsection{Frequency shift and superradiant speedup}
The resonant-beam-coupling geometry permits an analytical solution of \eqref{eq:dynamicsfinal}, which we briefly discuss here. 
The incidence angle $\theta_\mathrm{in}$ fixes the in-plane wavenumber to $k_\mathrm{in} = k_0 \cos \theta_\mathrm{in}$.
The excitation pulse can be written as $\Omega(x,t) = \Omega_0 \exp(i k_\mathrm{in} x) \Pi(t_\mathrm{ret})$, with $t_\mathrm{ret} = t - k_\mathrm{in} x / \omega_0$. 
Note that such an excitation pulse is made possible by the RBC geometry, where the incident plane wave is propagating in air/vacuum and therefore not attenuated. Other geometries will unavoidably introduce an off-resonant attenuation due to material absorption.

Taking the Fourier transform gives
\begin{equation}
\label{eq:omegagi}
\Omega(x,\omega) = \Omega_0 e^{i \tilde k_\mathrm{in} x}  \Pi(\omega),
\end{equation}
with $\tilde k_\mathrm{in} = (\omega + \omega_0)/c \cdot \cos \theta_\mathrm{in}$
Fourier-transforming \eqref{eq:dynamicsfinal} and inserting \eqref{eq:omegagi}, we obtain
\begin{equation}
\label{eq:dynamicsfourier}
    -i \pqty{\omega + i\gamma/2}\sigma_{ge}(x, \omega) + \frac{\gamma \zeta_m}{4 \Lambda_\mathrm{res}} \int_{-L}^x \dd{x'} e^{i \tilde k_m (x -x')} \sigma_{ge}\pqty{x',\omega}
    = i \Omega_0 \Pi(\omega) e^{i \tilde k_\mathrm{in} x} ,
\end{equation}
with $\tilde k_m = \nu_m (\omega + \omega_0)/c$.

The key to solve \eqref{eq:dynamicsfourier} is that, for $L \to \infty$, the right-hand side is an eigenfunction of the integral operator on the left-hand side. One readily obtains
\begin{equation}
    \sigma_{ge}(x, \omega) =   \frac{- \Omega_0 \Pi(\omega)}{\omega + i \gamma / 2 - \tilde \eta(\omega)}\exp(i \tilde k_\mathrm{in} x)
\end{equation}
where
\begin{equation}
    \tilde \eta(\omega) = \frac{\zeta_m \gamma}{4 \Lambda_\mathrm{res}} \frac{1}{(\omega_0 + \omega)/c} \frac{-1}{\nu_m - \cos \theta_\mathrm{in}}
\end{equation}
is the complex frequency shift.
Since $\omega \ll \omega_0$, we can neglect the frequency-dependency, setting $\eta := \tilde \eta(\omega = 0)$.
We write
\begin{equation}
    \eta = \frac{\gamma}{2} ~ \frac{\zeta_m \Lambda_m }{\Lambda_\mathrm{res}} ~ \frac{1}{2 \Lambda_m q_\theta - i}
\end{equation}
with $q_\theta = k_0 (\cos \theta_\mathrm{in} - \Re\{ \nu_m\})$.
Fourier-transforming the solution gives
\begin{equation}
\begin{split}
    \sigma_{ge}(x, t) &= \int \frac{\dd{\omega}}{2 \pi} e^{-i \omega t} \sigma(x, \omega) \\
    &= i \Omega_0 \Pi(\delta - i \gamma/2) \Theta(t_\mathrm{ret}) \exp[ik_\mathrm{in} x-(\gamma/2 -i \eta) t_\mathrm{ret}].
    \end{split}
\end{equation}
Finally, we approximate $\Omega_0 \Pi(\eta - i \gamma/2) \approx A$ to be the pulse area, taking into account that the frequency detuning is small compared to the bandwidth of the excitation pulse.

\subsection{The emitted field}
To compute the field emitted by the nuclei, we use \eqref{eq:bmicroscopic} and apply the same assumptions as before, in particular the asymptotic expansion of the Green's function.
We obtain
\begin{equation}
\label{eq:bmacroscopic0}
\begin{split}
    \vb B^+_\mathrm{nuc}(x, z, \omega) 
    &= \mu_0 \rho d \pqty{\frac{\omega+\omega_0}{c}}^2 \int  \tens{G}_{\mathrm{m},\mathrm{2d}}(x-x', z, z_0, \omega+\omega_0) \cdot \vb m_0\,\sigma_{ge}(x', \omega) \dd {x'} \\
    &\approx i k \mu_0 \rho d \frac{u_m(z) u_m(z_0)}{2 \nu_m} \vb m_0 \int_{-L}^0 e^{i \tilde k_m \abs{x-x'}} \sigma_{ge}(x', \omega) \dd{x'}.
\end{split}
\end{equation}
The constant factor can be brought into a more expressive form, so that
\begin{equation}
\label{eq:bmacroscopic1}
   \vb B^+_\mathrm{nuc}(x, z, \omega)  = - \vb B^+_0 \zeta_m \frac{u_m(z)}{u_m(z_0)} \frac{\gamma \tau}{4 \Lambda_\mathrm{res}} \frac{1}{i A} \int_{-L}^0 e^{i \tilde k_m \abs{x-x'}} \sigma_{ge}(x', \omega) \dd{x'},
\end{equation}
where $\tau = \Pi(\omega=0)$ is the length of the excitation pulse and the field amplitude is $\vb B_0^+ = \vb m_0 \Omega_0 \hbar / \abs{\vb m_0}^2$.
To compute the field inside the sample, we can again employ the forward scattering approximation or integrate \eqref{eq:bmacroscopic1} numerically. In the experiment, we measure the far-field pattern of the field at the back end of the waveguide ($x=0$). 

The integral can be readily evaluated analytically for the two cases discussed in the letter.
For the FC geometry we obtain
\begin{equation}
\label{eq:bmacroscopicfc}
   \vb B^+_\mathrm{nuc}(0, z, t)  = - \vb B^+_0 \frac{u_m(z)}{u_m(z_0)}\frac{\gamma \tau}{4} \bqty{\frac{\zeta_m L}{\Lambda_\mathrm{res}} \exp(i k_0 \nu_m L)} \exp(- \gamma t_\mathrm{ret}/2) 
    \frac{2 J_1\pqty{\sqrt{\gamma t_\mathrm{ret} \zeta_m L/\Lambda_\mathrm{res}}}}{\sqrt{\gamma t_\mathrm{ret} \zeta_m L/\Lambda_\mathrm{res}}}.
\end{equation}
For the RBC geometry, we obtain
\begin{equation}
\label{eq:bmacroscopicgi}
   \vb B^+_\mathrm{nuc}(0, z, t)  = - \vb B^+_0 \frac{u_m(z)}{u_m(z_0)}\frac{\gamma \tau}{4} \bqty{\frac{2 \zeta_m \Lambda_m}{\Lambda_\mathrm{res}} \frac{-i }{2 q_\theta \Lambda_m - i}} \exp\bqty{- (\gamma /2 - i \eta)t_\mathrm{ret}},
\end{equation}
retaining the frequency shift and superradiant decay. Here, the overall intensity decreases when the $\theta_\mathrm{in}$ is detuned from the mode resonance due to the wavelength mismatch.

\section{Samples and experiment}

\subsection{Layer design}
The waveguides were designed using the xwglib software described in Ref.\ \cite{Lohse_OE_2024}. 
The materials were chosen for minimal mode absorption while being suitable for sputter deposition.
The cladding thickness was set to \SI{30}{\nm}, which is thick enough to fully contain the resonant modes (effectively infinite) but still relatively thin to minimize roughness build-up.
Table \ref{tab:designpars} lists the model parameters computed for the two final layer designs, assuming perfectly sharp interfaces. 

\begin{table}[htb]
    \centering
    \begin{tabular}{S S S S S}
    \toprule
      {$m$} & {$\zeta_m$ (\num{e-2})} & {$1 - \Re\{\nu_m\}$} & {$\Im\{\nu_m\}$} & {$\Lambda_m$ (mm)} \\
    \midrule
    \multicolumn{5}{l}{FC} \\
        1 & 3.8 & 3.8e-6 & 2.8e-8 & 0.25 \\
        2 & 0   & 6.8e-6 & 6.8e-8 & 0.11 \\
    \midrule
    \multicolumn{5}{l}{RBC} \\
        1 & 2.1 & 2.9e-6 & 1.3e-8 & 0.51 \\
        2 & 0   & 3.9e-6 & 2.2e-8 & 0.31 \\
        3 & 2.3 & 5.9e-6 & 7.5e-8 & 0.09 \\
        4 & 0   & 8.2e-6 & 1.6e-7 & 0.04 \\
    \bottomrule
    \end{tabular}
    \caption{Model parameters for the resonant modes $m$ of the two waveguide designs: coupling coefficient $\zeta_m$, effective mode index $\nu_m$, and (off-resonance) mode attenuation length $\Lambda_m$ for the supported resonant modes.}
    \label{tab:designpars}
\end{table}

\subsection{Sample preparation}

The two samples were fabricated by magnetron sputter deposition, using an argon plasma at a pressure of \SI{5e-3}{\milli\bar} at a target-to-substrate distance of about \SI{7}{\cm}
with sputtering guns from the A300-XP series (AJA International, Massachusetts, USA).
The residual base pressure prior to deposition was typically less than \SI{4e-7}{\milli\bar}. Molybdenum and the \fefs-isotope-enriched iron (\SI{97}{\percent} enrichment) were deposited using targets with diameter \SI{3.81}{\cm} at \SI{12}{\watt} DC, and \ce{B_4C} using a target with diameter \SI{5.08}{\cm} at \SI{53}{\watt} RF power. The deposition rates were calibrated based on x-ray reflectivity measurements.

\begin{figure}[htb]
    \centering
    \includegraphics{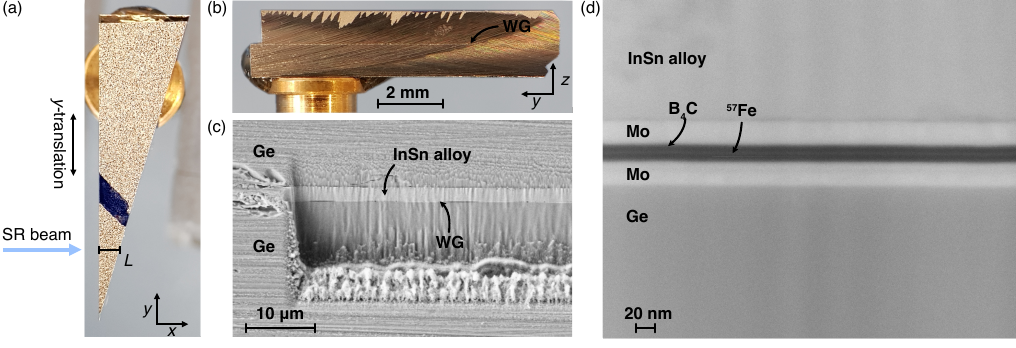}
    \caption{The FC waveguide assembly: (a) top view, showing the sample along the $y$-direction to change the propagation length $L$ of the x-rays in the WG; (b) Front-view; (c) the two wafers were bonded with an InSn alloy, visible as a horizontal stripe at medium magnification; (d) high-resolution view of the layer structure.}
    \label{fig:sample}
\end{figure}

To block the overilluminating part of the x-ray focus in the FC experiment, we bonded a second \SI{1}{\mm}-thick germanium wafer on top of the layer structure, as detailed in Ref.\ \cite{Krueger_JoSR_2012}.
Figure \ref{fig:sample}(a,b) show photographs of the final FC sample after bonding and cutting it to shape. 
To verify the layer structure of the FC sample, we acquired high-resolution cross-sectional SEM images. To that end, the layer structure was exposed with a Helios G4 focussed ion beam system (Thermo Fisher Scientific, Massachusetts, USA). The SEM images were acquired with an eLiNE  system (Raith GmbH, Germany).
Panels (c) and (d) show such images at different magnification. At highest magnification, the individual films are clearly distinguishable.

\subsection{Experimental setup}
\begin{figure}[htb]
    \centering
    \includegraphics{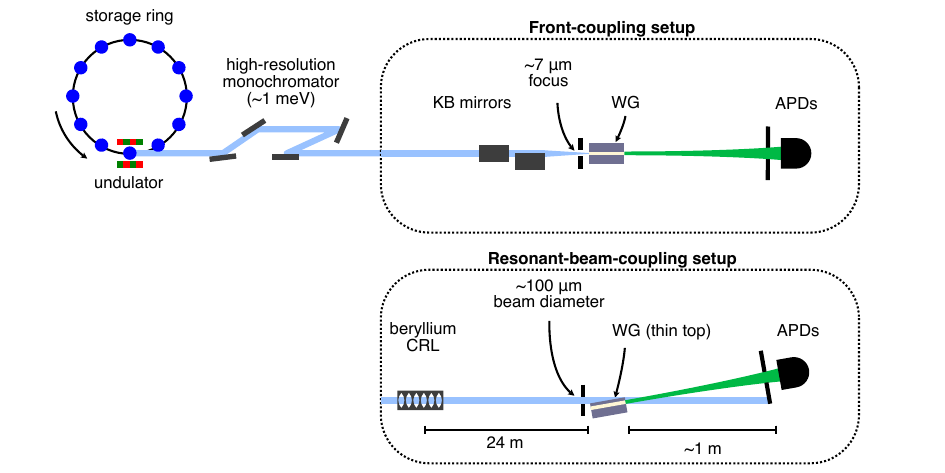}
    \caption{Experimental setups at the high resolution dynamics beamline P01 for realization of the front-coupling and the grazing-incidence geometries. Only the relevant components are sketched. In both experiments, a time-insensitive pixel detector was mounted \SI{3}{\metre} downstream of the WG (not shown). The APDs were moved out of the beam to acquire images with the pixel detector.}
    \label{fig:setup}
\end{figure}

Figure \ref{fig:setup} shows the setups of the two experiments performed at P01. The undulator radiation is monochromatized to about \SI{1}{\meV} by a high-resolution monochromator (HRM). In the front-coupling experiment, the beam was focussed by two mirrors in Kirkpatrick Baez geometry to about \SI{7}{\um} in focus diameter. A pair of slits was used to cut off the outer tails of the focus that were not blocked by the \SI{2}{\mm} thick waveguide assembly. A stack of four avalanche photo diodes (APDs), mounted on a translation stage, approximately \SI{1}{\metre} downstream of the waveguide was used to detect the photons exiting the waveguide.
In the resonant-beam-coupling experiment, the beam was mildly focussed by a beryllium compound refractive lens (CRL) down to about \SI{100}{\um} diameter. The CRL stack was placed immediately after the monochromator to maximize the focal distance and minimize angular divergence of the beam, which was about \ang{0.002}. The beam was cut in vertical direction to about \SI{20}{\um} using a slit placed immediately in front of the sample.
The angular acceptance of the APDs was adjusted with a pair of slits mounted in front of the APDs.
In both experiments, a time-insensitive pixel detector, a LAMBDA 750k (X-Spectrum GmbH, Hamburg, Germany) with \SI{55}{\um} pixel size, was placed about \SI{3}{\metre} downstream of the sample to measure the off-resonant (prompt) photons exiting the waveguide.

\begin{figure}[htb]
    \centering
    \includegraphics{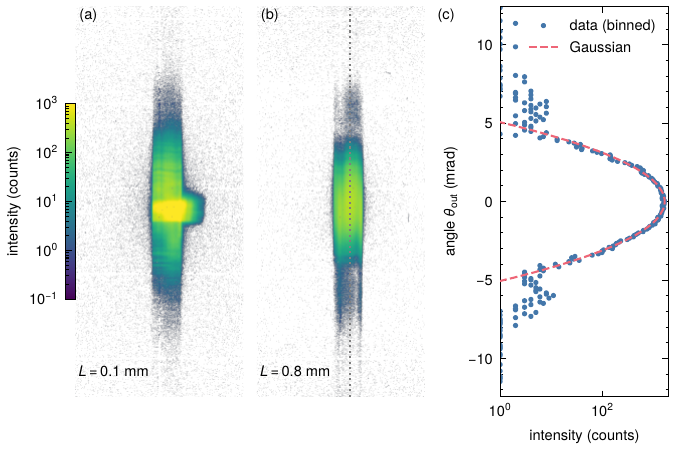}
    \caption{Off-resonant far-field diffraction patterns in the front-coupling geometry, acquired with time-insensitive pixel detector. (a,b) diffraction pattern for waveguide lengths $L = \SI{0.1}{\mm}$ and \SI{0.8}{\mm}, respectively. The primary beam is fully absorbed for the longer waveguide (b) but still visible in the shorter one (a). (c) Vertical profile [dotted line in (c)] through the diffraction pattern in logarithmic scale with a Gaussian fit that corresponds to a fully vertically coherent exit beam with FWHM \SI{19.8}{\nm}.} 
    \label{fig:farfield}
\end{figure}

Figure \ref{fig:farfield} shows far-field diffraction patterns, acquired with the LAMBDA 750k pixel detector in front-coupling geometry. Panels (a) and (b) illustrate how the primary beam is absorbed by the waveguide so that only the coherent guided mode is transmitted. The one-dimensional far-field pattern (c) is consistent with a fully transversally (in the vertical direction) coherent Gaussian mode with FWHM \SI{19.8}{\nm} at the end of the waveguide, as discussed in Ref.\ \cite{Osterhoff_NJoP_2011}.

\section{Data analysis}

\subsection{Approach}
The theory presented in the Letter captures the central features of the exciton dynamics and spontaneous photon emission and illustrates the qualitative behaviour. To obtain analytical solutions, we neglected the energetic sublevels of the nuclei, which will be slightly perturbed in the presence of hyperfine interactions. Even very weak hyperfine interactions affect the observable time-dependence and need to be taken into account in the analysis.
In principle, the theory can be readily extended to include the full hyperfine structure of the nuclei and then solved numerically \cite{Andrejic_PRA_2024}.

However, our theory shows that the time-dependence of the emitted fields is formally equivalent to that known from established experimental geometries.
In FC geometry, if only a single guided mode contributes, the time-dependence resembles NFS through a foil with rescaled thickness. In RBC geometry, the time-dependence is the same for the emission into the mode and into the direction of specular reflection---the well-known grazing incidence reflection geometry \cite{Roehlsberger_HI_1999}. This allows us to use available and tested software for simulation and data analysis of nuclear resonant scattering to model the experimentally observed temporal emissions. We emphasize that only the time-dependence is equivalent, whereas the spatial emission structure is clearly different.

\subsubsection{Hyperfine parameters}
The iron films in the samples are too thin to develop magnetic order so that we can neglect magnetic hyperfine splitting. However, variations in the electric field at the positions of the nuclei may cause inhomogeneous line broadening. An electric field gradient couples with the non-vanishing quadrupole moment of the excited state and may cause a splitting into two lines. Since the thin iron layer is likely amorphous, there is no overall directional dependence and we can take the isotropic average. We assumed a normal distribution for the line broadening, parameterized by its full width at half maximum $w_\mathrm{HF}$. The quadrupole splitting is parameterized by the distance between the split lines $s_\mathrm{HF}$.

\subsubsection{Data regression approach}
We used the Nexus \cite{LarsBocklage__2023} library to implement the forward model for the delayed intensity $f(t, \vec p)$ and inferred the model parameters $\vec p$ by numerically finding the maximum likelihood estimator (MLE), given the data (histogram of counts over time delay) and assuming Poisson statistics.

The log-likelihood for Poisson-distributed data $\vec t = (t_1, \dots, t_N)$ and $\vec y = (y_1, \dots, y_N)$ can be written as 
\begin{equation}
\label{eq:likelihood}
   l(\vec p) = - \sum_j \bqty{ f(t_j, \vec p) - y_j \cdot \log f(t_j, \vec p) }, 
\end{equation}
where here $y_j$ denote the number of counts in bin corresponding to time $t_j$. 
We computed the MLE by numerically optimizing \eqref{eq:likelihood}, $\vec p_\mathrm{MLE} = \operatorname{argmax}_{\vec p \in \mathcal{D}}\Bqty{-l(\vec p)}$ within a reasonable domain $\mathcal{D}$.

\subsection{Front coupling}
We modeled the FC waveguide as a thick slab/foil of \fefs\ (isotopically enriched to \SI{95}{\percent}, density \SI{7.874}{\gram\per\cubic\cm}) in a nuclear forward scattering geometry. The density and abundance are only relevant to relate the optical thickness $\xi$ to the geometrical thickness $d$ of the slab---they do not enter the model. With no magnetic hyperfine splitting and isotropically averaged hyperfine interactions, we optimized for the following 4 free parameters: intensity scale $I_0$, optical thickness $\xi$, line broadening FWHM $w_\mathrm{HF}$, and quadrupole splitting $s_\mathrm{HF}$. This was done for each each waveguide length independently. The results are listed in Tab.\ \ref{tab:hyperfine}.

\begin{table}[htb]
    \centering
    \begin{tabular}{l S S}
    \toprule
     Parameter & {FC} & {RBC} \\
    \midrule
     Broadening FWHM $w_\mathrm{HF}$ ($\gamma$) & 6(2) & 4 \\
     Quad. splitting $s_\mathrm{HF}$ ($\gamma$) & 6(1) & 7  \\
    \bottomrule
    \end{tabular}
    \caption{Hyperfine parameters inferred from the experimental data, assuming normally distributed broadening and quadrupole splitting of the transition lines. For FC, the parameters were extracted from each WG length independently. The number in parentheses gives the standard deviation. For RBC, the parameters were extracted from a single measurement in reflection geometry at $\theta_\mathrm{in} = \theta_\mathrm{out} = \ang{0.5}$.}
    \label{tab:hyperfine}
\end{table}

\subsection{Resonant beam coupling}
\subsubsection{Extraction of initial decay rate}
To extract the initial speedup of the emissions in RBC geometry, we assumed a simple exponential decay $f(t, \vec p) = p_0 \exp(-p_1 \gamma t)$ and inferred the optimal parameters (scale $p_0$ and speedup $p_1$) by numerically computing the MLE based on the data in the time interval $\SI{13}{\ns} \le t \le \SI{30}{\ns}$.

\subsubsection{Estimation of hyperfine parameters}
To estimate the hyperfine parameters ($w_\mathrm{HF}$ and $s_\mathrm{HF}$) of the RBC waveguide, we measured the delayed photons in reflection geometry at $\theta_\mathrm{in} = \theta_\mathrm{out} = \ang{0.5}$. This angle is far from the waveguide resonances to ensure that the time-dependence (up to a global scaling) is to a good approximation unaffected by the waveguide structure, yet still sensitive to the hyperfine interactions. 

We modeled the experiment numerically, using ideal geometrical parameters (design layer thicknesses and densities, perfectly sharp interfaces). Similar to the FC geometry, we optimized for the following 3 free parameters: 
intensity scale $I_0$, line broadening FWHM $w_\mathrm{HF}$, and quadrupole splitting $s_\mathrm{HF}$. The results are also listed in Tab.\ \ref{tab:hyperfine}.

\subsubsection{Full simulation of speedup}

\begin{figure}[htb]
    \centering
    \includegraphics{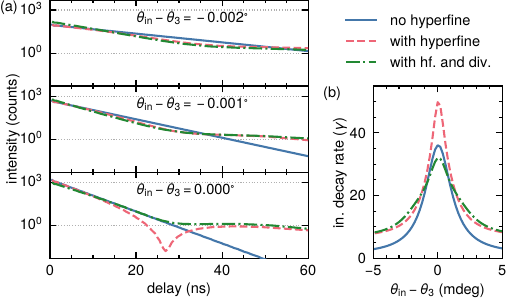}
    \caption{Simulated emission in reflection geometry ($\theta_\mathrm{in} = \theta_\mathrm{out}$) to describe the time-dependence in RBC excitation. (a) Time-dependence and (b) speedup. Without hyperfine splitting (blue), the simulation is in perfect agreement (up to global scaling) with the predictions of the simple theory. The additional beating due to the hyperfine interactions (green) results in increased (apparent) speedup. Close to resonance, this effect is partially reversed by the angular divergence of \SI{2.1}{mdeg} FWHM (orange), which is assumed to be uniform and incoherent.}
    \label{fig:simulation}
\end{figure}

To explain the values of the speedup in the RBC geometry, we simulated the emission from the layer system as a function of time in reflection geometry, $\theta_\mathrm{out} = \theta_\mathrm{in}$. To that end, we exploit that the time-dependence is the same for the emissions out of the back end of the WG and for the field that evanescently couples through the top cladding and is emitted into the direction of specular reflection. We used Nexus \cite{LarsBocklage__2023} to simulate the time-dependent field in reflection geometry, assuming an infinitely extended layer system. Note, however, that the angular dependence is different between the two geometries, so that the simulated intensity scale is meaningless.
The results, without hyperfine interactions, with hyperfine interactions (extracted from independent data as detailed above), and including a uniform angular divergence of \SI{2.1}{mdeg} (FWHM), are presented in Fig.\ \ref{fig:simulation}(a). 
The angular divergence is modeled as an incoherent average with uniformly distributed angles (arithmetic mean).

From these simulated curves, we extracted the initial decay rate, using the same procedure as for the experimental data. The results are shown and discussed in Fig.\ \ref{fig:simulation}(b). The simulation without hyperfine splitting, on the one hand, accurately matches the simple single-mode model. The full simulation including hyperfine splitting and divergence, on the other hand, explains the experimental data  (see main text). We conclude that the deviation from the simple model is indeed mainly caused by residual hyperfine splitting and beam divergence.

\bibliography{references}